# Computer simulation of a-Si/c-Si heterojunction solar cell with high conversion efficiency

Chen AQing, Zhu KaiGui*

Department of Physics, Beihang University, Beijing 100191, P. R. China.

**Abstract:** The p-type amorphous/ n-type crystalline silicon ($P^+$ a-Si/$N^+$ c-Si) heterojunction was simulated for developing the solar cells with high conversion efficiency and low cost. The characteristic of such cells with different work function of transparent conductive oxide (TCO) were calculated. The energy band structure, quantum efficiency and electric field are analyzed in detail to understand the mechanism of the heterojunction cell. Our results show that the a-Si/c-Si heterojunction is hypersensitive to the TCO work function, and the TCO work function should be large enough in order to achieve high conversion efficiency for $P^+$ a-Si:H/$N^+$ c-Si solar cell. With the optimized parameters set, the $P^+$ a-Si:H/$N^+$ c-Si solar cell reaches a high efficiency ($\eta$) up to 21.849% (FF: 0.866, $V_{OC}$: 0.861 V, $J_{SC}$: 29.32 mA/cm$^2$).

1. **Introduction**

Silicon solar cell has been continuously paid attention as one of the important ways of solar energy. A silicon solar cell is a typical photovoltaic cell. Zhao (1998) et al fabricated the poly-silicon and monocrystalline silicon solar with conversion efficiencies of 19.8% and 24.4%, respectively. The amorphous silicon solar cell of multilayered p-i-n unit cell structure with a high open voltage of 2.0 V has been developed in early 1979 (Hamakawa et al., 1979). The thin solar cells fabricated by Michael Reuter (2009) et al have the conversion efficiency of 17.0%. In order to improve the conversion efficiency, the front surface of cells was textured (Muller et al. 2004). In some cases, additional textured photonic crystal and backside reflector (Zeng et al., 2008) were fabricated at the back surface to facilitate efficient light trapping.

Heterojunction solar cells studied so far include a-Si ITO/p-i-n heteroface solar cells (Okamoto et al., 1980), amorphous silicon oxide/microcrystalline silicon double-junction solar cells (Sriprapha et al., 2011; Yuan et al., 2010), a-si/poly-silicon solar cell (Chen and Shao, 2011), a-Si:H/c-Si heterojunction solar cells (Hernández-Como and Morales-Acevedo, 2010; Fuhs et al., 2006; Tsunomura et al., 2009; Tanaka et al., 2003). Heterojunction solar cells with textured surface have a high conversion efficiency of over 22% for laboratory cells and over 19.5% for mass production cells (Taira et al., 2007). However, this heterojunction solar cell has a complicated structure with a very thin intrinsic a-Si:H(i) layer inserted between p-type a-Si:H and Czochralski (CZ) crystalline silicon (c-Si). The cost of this heterojunction cell can be cut down greatly owing to the fact that a-Si film can be prepared at low temperature by hot-wire chemical vapor deposition (Villara et al., 2009). Numerical simulation results suggested that the conversion efficiency of a-Si:H/c-Si solar cell decreased as the increase of interfacial density of states (Hernández-Como and Morales-Acevedo, 2010). Those cells have good performance, but they had complicated structure leading to a high cost. With the purposes of reducing the cost further and promoting the performance of heterojunction solar cells, we designed a simple cell structure, namely, p-type a-Si was deposited on one side of n-type CZ c-Si forming a p-n junction. By simulation of this device with optimized parameters, we found that this solar cell has a high performance with conversion efficiency of 21.849 %, fill factor (FF) of 0.866, open circuit voltage ($V_{OC}$) of 0.861 V, and short circuit current density ($J_{SC}$) of 29.32 mA/cm$^2$. Compared to present products and reported work (Hernández-Como and Morales-Acevedo, 2010; Fuhs et al., 2006; Tsunomura et al. 2009), the as-designed heterojunction solar cell has advantages in terms of both cost and performance. We calculated the work function of TCO, and discussed the reasons for the high performance of this kind of heterojunction solar cell in terms of band structure, quantum efficiency and electric field. The result presents a significant improvement of solar cells in terms of high conversion efficiency and $V_{OC}$.

## 2. Structure of a-Si:H/c-Si heterojunction solar cell

Figure 1 shows the structure of P$^+$ a-Si:H/ N$^+$ c-Si heterojunction solar cell. The P$^+$ a-Si:H layer as the emitter is deposited on the n-type CZ c-Si wafer used as an absorption layer, forming a p-n heterojunction. Transparent conductive oxide (TCO) and aluminum film are deposited on the front of P$^+$ a-Si:H layer and the back of the n-type CZ c-Si wafer, respectively. Sun light travels from the top to the bottom. The AMPS-1D, an effective and powerful software for analyzing and designing transport in microelectronic and photonic structures (Zhu et al., 1999), was employed to

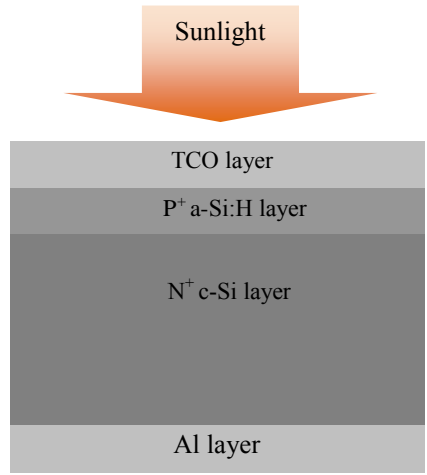

Figure 1 the structure of P$^+$ a-Si:H/ N+c-Si heterojunction solar cell simulate this cell. AMPS-1D has two pictures, namely, the Lifetime and DOS pictures. In the DOS picture the details of recombination traffics, trapping and the charge state of the defects (and the effects of this charge on the electric field variation across a structure) are fully explained. The DOS picture, hence, was employed to simulate the P$^+$ a-Si/N$^+$ c-silicon solar cell structure. The data of a-Si:H is obtained from the compendium of parameters of AMPS-1D. The mobilities of electron and hole were set as 100 and 50 cm2 V$^{-1}$ s$^{-1}$, respectively, according to the reference (Sze, 2007). Both the doping concentrations of acceptors ($N_A$) and donors ($N_D$) are $1.0 \times 10^{19}$ cm$^{-3}$. The p type a-Si with $N_A = 1.0 \times 10^{19}$ has heavy doping effects (Markvart and Castañer, 2005), namely, impurity band is formed and gradually merges with the parent band and eventually gives rise to a tail of the localized states. Such tail could extend further into the gap due to overlap with the localized gap states (Fonash, 2010). As a result, p type a-Si has a broaden absorption edge and therefore a stronger optical absorption than intrinsic a-Si. Considering the optical absorption of p type a-Si is strengthen mainly in the infrared edge range, we assume the optical absorption of p type a-Si layer as the same as that of the intrinsic layer for 330 - 900 nm range which is well known (Searle, 1998). For the simulation illumination condition, we used the AM 1.5 spectrum normalized to 100mW/cm$^2$. In the following calculations, the thickness of P$^+$ a-Si:H and N$^+$ c-Si were kept as 10 nm and 300 μm, repectively.

## 3. Results and discussion

The work function of transparent conductive oxide (TCO) has great impacts on the P$^+$ a-Si:H/N$^+$ c-Si heterojunction solar cell (Delahoy et al., 2008; Zhao et al., 2008). The work function of TCO ($W_{TCO}$) is generally about 4.7eV (Chkoda et al., 2000; Hashimoto et al., 2001). ATO, AZO, and ITO have relatively low work functions, which are 3.8 - 5.2 eV, 3.1 - 4.5 eV, 3.6 - 5.3 eV, respectively (Klein et al., 2010). Comparatively, zinc-indium-tin oxide (ZITO), gallium-indium-tin oxide (GITO) and gallium-indium oxide (GIO) have relatively high work functions of 5.2 - 6.1 eV as well as high optical transparencies (>90%) (Marks et al., 2002). It is very important to determine the proper work function of TCO for designing high conversion efficiency P$^+$ a-Si:H/N$^+$ c-Si heterojunction solar cell. So, it is necessary to calculate the influence of work function on the heterojunction solar cell. The simulated performance of P$^+$ a-Si:H/N$^+$ c-Si solar cell as a function of work function is shown in the figure 2. The thickness of N$^+$ c-Si layer is 300μm and both $N_D$ and $N_A$ are $1.0 \times 10^{19}$ cm$^{-3}$ as mentioned above. It is easy to note that when $W_{TCO}$ is larger than 4.3eV, the work function has little impacts on the $J_{sc}$. While the conversion efficiency, FF and $V_{OC}$ increase rapidly with $W_{TCO}$, and then become saturated at $W_{TCO}$ = 5.1 eV.

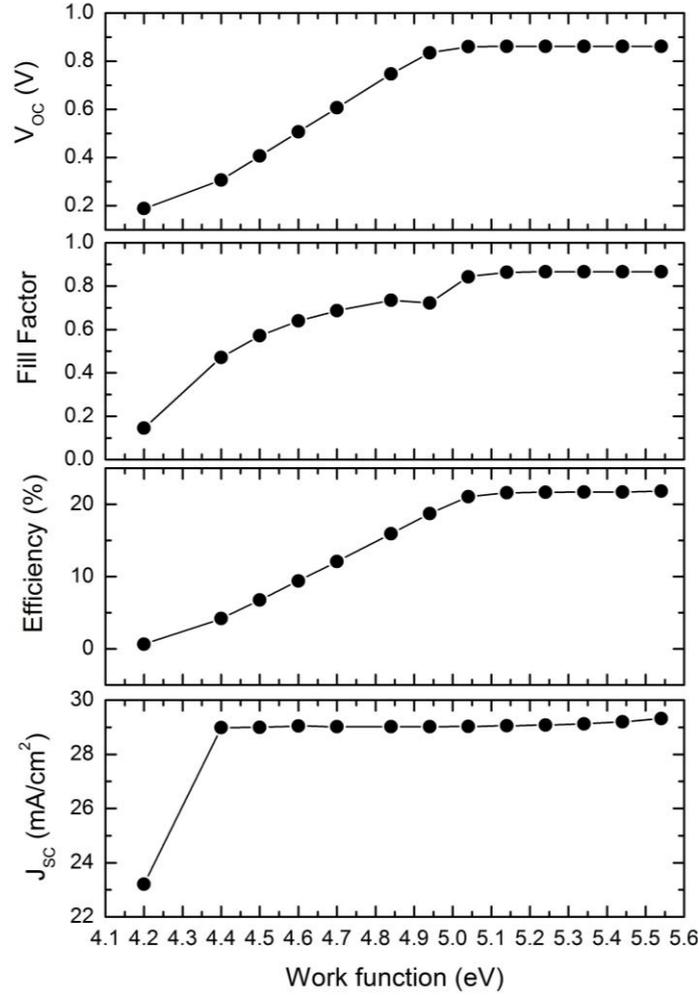

Figure 2 simulated performance of $P^+$ a-Si:H/ $N^+$c-Si heterojunction solar cell as a function of TCO work function with the thickness of $N^+$ c-Si of 300μm, $N_D = N_A = 1.0 \times 10^{19}$ cm$^{-3}$

We can achieve the maximum efficiency about 21.849% at $W_{TCO}$ = 5.54 eV. It is obvious that at $W_{TCO}$ = 4.2 eV, the cell has a poor performance with η = 0.638%, FF = 0.146, $V_{OC}$ = 0.189 V, $J_{SC}$ = 23.212 mA/cm$^2$, while at $W_{TCO}$ = 5.54 eV, the cell has a good performance with η = 21.849%, FF = 0.866, $V_{OC}$ = 0.861 V, $J_{SC}$ = 29.32 mA/cm$^2$.

In order to understand the effects of work function on the $P^+$ a-Si:H/$N^+$ c-Si heterjunction solar cell in depth, it is necessary to take into account the energy band structure and the built-in electric field. The energy band structure of the cell with TCO work function of 4.2eV and 5.54eV are shown in figure 3 (a) and (b), respectively. The grey region in figure 3 represents the $P^+$ a-Si:H layer. $\chi_1$ is the electron affinity. $V_B$ and $V_b$ are the build-in potential. $\Delta_v$ is the affinity step (Fonash, 2010). The build-in potential $V_B$ is always equal to the work function difference. Besides, TCO/P$^+$ a-Si:H Schottky contact and P$^+$ a-Si:H/N$^+$ c-Si p-n junction locate on each side of the P$^+$ a-Si:H layer regarded as the emitter. When $W_{TCO}$ is low, the Fermi energy level of TCO is higher than that of P$^+$ a-Si:H layer resulting in the fact that electrons flow out of TCO layer into P$^+$ a-Si:H layer. So both conduction and valence band bend downward at the interface between TCO and P$^+$ a-Si:H layer, as shown in Figure 3 (a), which leads to a built-in potential $V_b$ with the opposite direction to $V_B$, that of P$^+$ a-Si:H/N$^+$ c-Si p-n junction. Thus, the barrier of TCO/P$^+$ a-Si:H Schottky contact hinders the holes from transporting to the front electrode resulting in the great decrease of the conversion efficiency. In contrast, when $W_{TCO}$ is high, the Fermi energy level of TCO layer is equal to or lower than that of P$^+$ a-Si:H layer resulting in the fact that electrons from P$^+$ a-Si:H layer flow into the TCO layer. Both conduction and valence band bend upward at the interface between TCO and P$^+$ a-Si:H layer, which leads to a built-in potential with the same direction to that of P$^+$ a-Si:H/N$^+$ c-Si p-n junction. As a results, the barrier of TCO/P$^+$ a-Si:H Schottky is lessened even to be negligible, as shown in figure 3 (b). The photogenerated holes, consequently, can arrive at the electrode more easily so that the conversion efficiency can be enhanced.

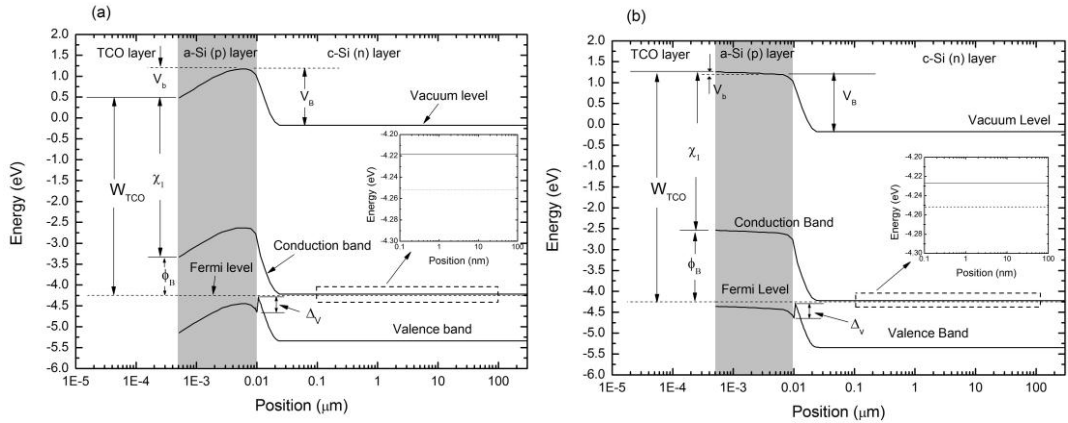

Figure 3 (a) and (b) show the energy band structure of the cell with WTCO =4.2 and 5.54 eV, respectively.

Furthermore, the quantum efficiency, the ratio of the number of electrons in the external circuit produced by an incident photon of a given wavelength (Markvart and Castañer, 2005), is analyzed to explain the outstanding performance of this cell, as shown in Figure . As we know, the photogenerated current $I_{ph}$ is derived from the

equation

$$I_{ph} = q\int_{(\lambda)} \phi(\lambda)\{1-R(\lambda)\}QE(\lambda)d\lambda \quad (3)$$

Where $\phi(\lambda)$ is the photon flux incident on the cell at wavelength $\lambda$; $QE(\lambda)$ is the quantum efficiency; $R(\lambda)$ is the reflection coefficient from the top surface. The integration is carried out over all the range of wavelength $\lambda$ of light absorbed by the solar cell. It is seen in Figure that the high $W_{TCO}$ value of 5.54 eV can improve the quantum efficiency in the whole range of spectrum wavelength from 330 nm to 900 nm. According to the equation (3), this will lead to the higher $I_{ph}$ value at $W_{TCO}$ = 5.54eV than the value at $W_{TCO}$ = 4.2eV. On the other hand, the $V_{OC}$ is given as

$$V_{OC} = \frac{k_B T}{q}\ln\left(1+\frac{I_{ph}}{I_0}\right) \quad (4)$$

From the above equation, the $V_{OC}$ increase logarithmically with the photogenerated current $I_{ph}$ and reaches the maximum value equal to the contact potential. Hence, when $W_{TCO}$ = 5.54 eV, the maximal $V_{OC}=V_D \approx V_B$ (see figure 3). The conversion efficiency $\eta$ is written as (Sze, 2007)

$$\eta = \frac{I_{ph}}{q}\left(\frac{h\nu}{P_{opt}}\right) \quad (5)$$

Where the $P_{opt}$ is the optical power. So the $\eta$ also increases with the TCO work function according to equation (3), (5) and the distribution of quantum efficiency vs wavelength.

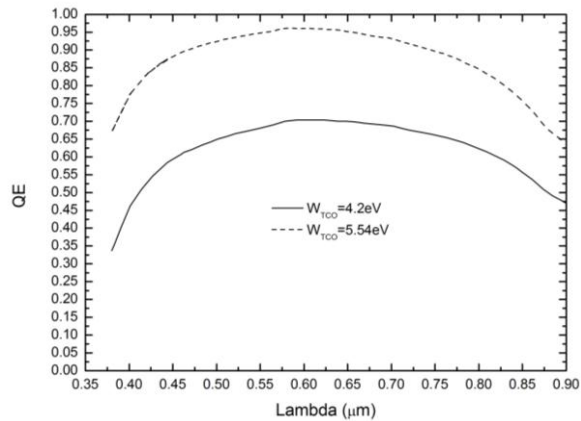

Figure 4 the quantum efficiency of the cell for TCO work function of 4.2eV and

5.54eV vs. wavelength.

Different concentration of carries at the interface between TCO and $P^+$ a-Si:H layer will lead to different built-in electric filed. Figure 5 shows the distribution of built-in electronic filed as a function of work function of TCO. We can see that the electric filed decreases with the increase of the TCO work function in grey region, namely, the built-in potential decreases gradually. Nevertheless, in the region of $N^+$ c-Si layer the electric field keeps unchanged. In the grey region with $W_{TCO} < 5.14eV$, the electric filed is positive near the interface, which hinders holes from reaching the front electrode. When $W_{TCO}$ is 5.54eV, the electric field is negative in the whole grey region. This electric field gives the effective force on holes to sweep them to the left region in the figure 3, which is very benefit for holes to transport through the grey region.

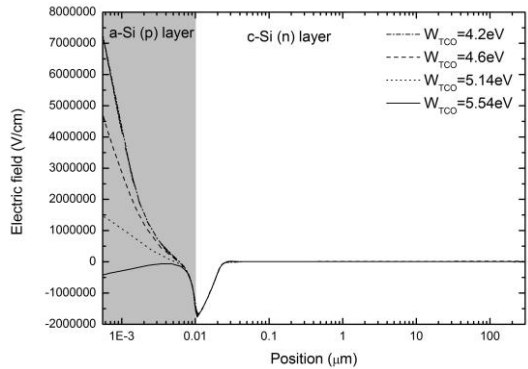

Figure 5 the distribution of electric filed with various work function of TCO

By above analysis it is obtained that the work function of TCO should be large enough to achieve higher conversion efficiency for $P^+$ a-Si:H/$N^+$ c-Si solar cell. The result can help the development of higher conversion efficiency and lower cost solar cells.

## 4. Conclusion

The main effective factor, work function of TCO, responsible for the outstanding performance of such a $P^+$ a-Si:H/$N^+$ c-Si solar cell is discussed, and it is found that the TCO work function should be large enough to achieve high conversion efficiency. Energy band structure, quantum efficiency and distribution of electrical field are analyzed to understand mechanism of the high performance. By numerical simulation, we designed a prospective $P^+$ a-Si:H/$N^+$ c-Si solar cell of high performance with high efficiency of above 21.894%, high fill factor of 0.866 and high $V_{oc}$ of 0.861 V. The simulation result may benefit the exploration of the solar cells with high conversion efficiency and low cost.

## Acknowledgements:

The authors gratefully acknowledge Professor S Fonash of the Pennsylvania State University and the Electric Power Research Institute for providing the AMPS-1D program used in the simulations.## References

Chen, A.Q., Shao, Q.Y., 2011. Simulation of high conversion efficiency and open-circuit voltages of α-si/poly-silicon solar cell. Sci China Phys Mech Astron 54, 1466-1470.

Chkoda, L., Heske, C., Sokolowski, M., Umbach, E., 2000. Improved band alignment for hole injection by an interfacial layer in organic light emitting devices. Appl. Phys. Lett. 77, 1093-1095.

Delahoy, A.E., Stavrides, A.P., Patel, A.M., et al., 2008. Influence of TCO Type on the Performance of Amorphous Silicon Solar Cells. Proc. of SPIE 7045, 704506.

Fonash, S. J., 2010. Solar cell device physics. Academic Press, Inc., New York,

Fuhs, W., Korte, L., Schmidt, M., 2006. Heterojunctions of hydrogenated amorphous silicon and monocrystalline silicon. Journal of Optoelectronics and advanced materials 8, 1989-1995.

Okamoto, H., Nitta, Y., Yamaguchi, T., Hamakawa,Y., 1980. Device physics and design of a-Si ITO/p-i-n heteroface solar cells, Solar Energy Materials 2, 313-325.

Hamakawa, Y., Okamoto, H., Nitta, Y., 1979. A new type of amorphous silicon photovoltaic cell generating more than 2.0 V. Applied Physics Letters 35, 187 - 189.

Hashimoto, Y., Hamagaki, M., and Sakakibara, T., 2001. Effect of Ionization Potential of Hole Transport Layer on Device Characteristics of Organic Light Emitting Diode with Oxygen Plasma Treated Indium Tin Oxide. Jpn. J. Appl. Phys. 40, 4720-4725.

Hernández-Como, N., Morales-Acevedo, A., 2010. Simulation of hetero-junction silicon solar cells with AMPS-1D. Solar Energy Materials & Solar Cells 94, 62-67.9